\documentclass[10pt,conference,letterpaper]{IEEEtran}

\usepackage{cite}
\usepackage{amsthm}
\usepackage{amsmath,graphicx}
\usepackage{amssymb}
\usepackage{amsfonts}
\usepackage{graphicx,subfigure}
\usepackage{fancyhdr}  %
\usepackage{cases}
\usepackage{extarrows}
\usepackage{algorithm}
\usepackage{algorithmic}
\usepackage{multirow,tabularx}
\usepackage{mathtools}
\usepackage{xcolor}
\usepackage[english]{babel}
\usepackage{caption}
\usepackage{bm}
\usepackage{graphicx}
\usepackage{epstopdf}
\usepackage{cuted}
\usepackage{stfloats}

\theoremstyle{definition}

\newtheorem{theorem}{Theorem}
\newtheorem{lemma}{Lemma}

\newtheorem{proposition}{Proposition}

\def\proof{\noindent\hspace{2em}{\itshape Proof: }}
\def\endproof{\hspace*{\fill}~$\square$\par\endtrivlist\unskip}

\makeatletter
\renewcommand{\maketag@@@}[1]{\hbox{\m@th\normalsize\normalfont#1}}%
\makeatother

\begin{document}
	\IEEEoverridecommandlockouts
	\title{RIS-assisted Coverage Enhancement in mmWave Integrated Sensing and Communication Networks}

	\author{
		\IEEEauthorblockN{
			Xu Gan\IEEEauthorrefmark{1}\IEEEauthorrefmark{2},
			Chongwen Huang\IEEEauthorrefmark{1}\IEEEauthorrefmark{2},
			Zhaohui Yang\IEEEauthorrefmark{1},
			Xiaoming Chen\IEEEauthorrefmark{1},
			Faouzi Bader\IEEEauthorrefmark{3},
			Zhaoyang Zhang\IEEEauthorrefmark{1},\\
			Chau Yuen\IEEEauthorrefmark{4},
			Yong Liang Guan\IEEEauthorrefmark{4}, and
			M\'{e}rouane Debbah\IEEEauthorrefmark{5}}
		\IEEEauthorblockA{\IEEEauthorrefmark{1}College of Information Science and Electronic Engineering, Zhejiang University, 310027, Hangzhou, China}
		\IEEEauthorblockA{\IEEEauthorrefmark{2}State Key Laboratory of Integrated Service Networks, Xidian University, 710071, Xi'an, China}
		\IEEEauthorblockA{\IEEEauthorrefmark{3}Technology Innovation Institute, 9639 Masdar City, Abu Dhabi, UAE}
		\IEEEauthorblockA{\IEEEauthorrefmark{4}School of Electrical and Electronics Engineering, Nanyang Technological University, 639798, Singapore}
		\IEEEauthorblockA{\IEEEauthorrefmark{5}Khalifa University of Science and Technology, P O Box 127788, Abu Dhabi, UAE}

		\thanks{The work was supported by the China National Key R\&D Program under Grant 2021YFA1000500, 2023YFB2904800 and 2023YFB2904804, National Natural Science Foundation of China under Grant 62331023, 62101492, 62394290, 62394292, 62231009 and U20A20158, Zhejiang Provincial Natural Science Foundation of China under Grant LR22F010002, Zhejiang Provincial Science and Technology Plan Project under Grant 2024C01033, and Zhejiang University Global Partnership Fund.  The Prof. C. Yuen was supported by MOE Tier 2 Award number MOE-T2EP50220-0019.}
		\vspace{-30pt}
	}

	\maketitle
	
	\begin{abstract}
		Integrated sensing and communication (ISAC) has emerged as a promising technology to facilitate high-rate communications and super-resolution sensing, particularly operating in the millimeter wave (mmWave) band. However, the vulnerability of mmWave signals to blockages severely impairs ISAC capabilities and coverage. To tackle this, an efficient and low-cost solution is to deploy distributed reconfigurable intelligent surfaces (RISs) to construct virtual links between the base stations (BSs) and users in a controllable fashion. In this paper, we investigate the generalized RIS-assisted mmWave ISAC networks considering the blockage effect, and examine the beneficial impact of RISs on the coverage rate utilizing stochastic geometry. Specifically, taking into account the coupling effect of ISAC dual functions within the same network topology, we derive the conditional coverage probability of ISAC performance for two association cases, based on the proposed beam pattern model and user association policies. Then, the marginal coverage rate is calculated by combining these two cases through the distance-dependent thinning method. Simulation results verify the accuracy of derived theoretical formulations and provide valuable guidelines for the practical network deployment. Specifically, our results indicate the superiority of the RIS deployment with the density of 40 km${}^{-2}$ BSs, and that the joint coverage rate of ISAC performance exhibits potential growth from $67.1\%$ to $92.2\%$ with the deployment of RISs.

	\end{abstract}

	\section{Introduction}
	Integrated sensing and communication (ISAC) has been regarded as one most promising scenario for the sixth-generation (6G) networks, as it enables the capability to incorporate dual functions into the same wireless system to provide environment-aware services and improve resource utilization\cite{SG_ISAC4,RIS-ISAC2,ISAC1}. Capitalizing on the abundant bandwidth resources and beamforming technologies, the millimeter wave (mmWave) band can facilitate ISAC systems to simultaneously implement high-rate communications and super-resolution sensing services. It opens up revolutionary opportunities for supporting emerging applications, such as extended reality, metaverse and autonomous vehicles \cite{autonomous}, etc. There has been intensive research to explore the huge potential of mmWave ISAC systems \cite{mmISAC1,mmISAC2}, which is expected to benefit of the 6G communications.

	However, the inherent high propagation loss in the mmWave band may severely weaken ISAC signals by creating network blind zones that impair both communication and sensing performance. Fortunately, the advent of the reconfigurable intelligent surface (RIS) provides an attractive solution to tackle this challenge by manipulating the wireless propagation environment\cite{RIS1,RIS2,RIS3,RIS4}. Specifically, RIS is a programmable electromagnetic metasurface consisting of a large number of low-cost reflective elements. By virtue of its ability to adjust the phase and amplitude of incident signals, RIS is capable of establishing strong virtual line-of-sight (LoS) links to boost received signal strength. Due to the above benefits of incorporating RIS into the mmWave ISAC systems, it has gained widespread attention and interests in the existing literatures\cite{RIS-ISAC1,RIS-ISAC2,RIS-ISAC3}. Specifically, \cite{RIS-ISAC2} developed a transmission scheme to superimpose sensing pilots onto data information, which can achieve centimeter-level localization while sacrificing negligible communication spectral efficiency. In \cite{RIS-ISAC1}, the RIS-aided ISAC framework to extract the inherent location information in the received communication signals from a blind-zone user was proposed. Indeed, these works delicately investigated the utilization of RIS to improve the ISAC performance. Nevertheless, how much performance gain or coverage improvement RIS can offer in ISAC networks with blockages remains an unaddressed question.

	To answer this fundamental question, we can resort to stochastic geometry for modelling RIS-assisted ISAC networks. Specifically, stochastic geometry\cite{SG1} is a powerful tool to consider all possible locations of users, BSs and RISs, thereby facilitating the generalized derivation of performance metrics at a network level. However, existing stochastic geometry research primarily focuses on RIS-assisted communication networks\cite{SG_RIS1,SG_RIS2,SG_RIS3,SG_RIS4}. For instance, \cite{SG_RIS1} studied the effect of large-scale deployment of RISs on the ratio of the blind spots to the total area. \cite{SG_RIS2} provided communication capacity and energy efficiency of realistic user association and two-hop RIS model. Most recently, \cite{SG_RIS4} investigated the coverage probability and ergodic rate in RIS-assisted mmWave communication systems under single-cell and multi-cell scenarios.
	
	Based on the above discussion, it is crucial to derive fundamental limits for the coverage rate gain from RISs in ISAC networks. In this paper, we provide theoretical results for the coverage rate of RIS-assisted ISAC networks to quantify the coverage enhancement attributed to RISs. Specifically, the location distributions of users, BSs, RISs and blockages are modelled by the broadly applied stochastic geometry. Besides, leveraging the properties of mmWave beamforming technologies enabled by the large-scale antenna array, the beam pattern at the BS and the RIS can be modelled in a sector-based probability manner according to the width and amplitude of the main beam. Based on the proposed framework, we develop the user association policies and derive the conditional coverage rate under two association cases. Finally, we can calculate the marginal coverage rate using the distance-dependent thinning method. Simulation results indicate that the joint coverage rate of ISAC performance can grow from $67.1\%$ to $92.2\%$ with the deployment of RISs. These findings also serve as essential instruction and guidelines for the practical BS and RIS deployment. Specifically under the blockage-heavy scenario, the superiority of the RIS deployment with the density of 40 km${}^{-2}$ BSs is evident.

	\section{System Model}\label{model}

	\subsection{Network Model}
	In this work, we considered RIS-assisted ISAC networks, where each BS is equipped with $M_T$ transmit antennas and $M_R$ received antennas, and the locations of BS follow a homogeneous Poisson point process (PPP), denoted by $\Phi_B$ with density $\lambda_B$. We focus on a general ISAC scenario where the BS transmit signals to communicate with its association user while simultaneously extracting the location or posture information of this user from the reflected echoes. Besides, the locations of users and blockages are modelled as PPPs $\Phi_U$ with density $\lambda_U$ and $\Phi_L$ with density $\lambda_L$, respectively. These blockages can obstruct the LoS paths, and thus the LoS probability of the receiver at the distance $r$ can be derived in \cite{LoS_pro} as $P_{LoS}(r) = \exp(-\beta r)$, where $\beta = \frac{2 \lambda_L \mathbb{E}[L]}{\pi}$ and $\mathbb{E}[L]$ is the average length of the blockages. To enhance the ISAC performance, $N_R$-element RISs are distributely deployed as a PPP $\Phi_R$ with density $\lambda_R$. Then, the density of the LoS BS $\Phi_B^L$ and LoS RIS $\Phi_R^L$ with respect to the typical user can be derived in the following \emph{Proposition 1}.
	
	\begin{proposition}
		(Distance-Dependent Thinning of homogeneous PPP \cite{SG_RIS1}) When performing the distance-dependent thinning with probability $1-g(r)$ on a homogeneous PPP $\Phi$ with density $\lambda$, an inhomogeneous PPP $\tilde{\Phi}$ with density $\tilde{\lambda}(r) = g(r) \lambda$ will be obtained.
	\end{proposition} 
	
	As such, the densities of $\Phi_B^L$ and $\Phi_R^L$ can be written as $\lambda_B^L(r)=P_{LoS}(r)\lambda_B$ and $\lambda_R^L(\xi)=P_{LoS}(\xi)\lambda_R$, where $r$ and $\xi$ denote the distance from the typical user to the BS and the RIS, respectively. Furthermore, the density of virtual LoS (VLoS) BSs $\Phi_B^V$ is the multiplication of the NLoS BS density by the probability of the presence of a LoS RIS, i.e., $\lambda_B^V(r) = \lambda_B^N(r) \left( 1- P_{void}^{R,L}\right)$, where $P_{void}^{R,L} = \exp \left( -2\pi \int_0^{\infty} \lambda_R^L(r) r \mathrm{d} r \right)$.
	
	\vspace{-1mm}
	\subsection{Channel Model}
	We focus on two types of BS-user links, with the first being the LoS path and the second one being the cascaded path assisted by an RIS in case of blockages.
	
	\subsubsection{The direct channel}
	The LoS channel model between the $i$-th BS and the typical user can be expressed as $\mathbf{h}_i^{d} = \chi_d(d^{B-U}_i) \xi_i^d \mathbf{a}_{M_T}(\theta_{D,h_i^d})$, with $\mathbf{a}_M(\theta) = \left[1,e^{j2\pi\frac{d_{M} }{\lambda}\sin(\theta)},..., e^{j2\pi(M-1)\frac{d_{M} }{\lambda}\sin(\theta)} \right]$, where the path loss gain $\chi_d(d^{B-U}_i)=C_0 (d^{B-U}_i)^{-\alpha}$, $C_0$ is the path loss at the reference $1$ m, $d^{B-U}_i$ and $\xi_i^d$ are the distance and small-scale fading channel gain between the $i$-th BS and the user with $|\xi_i^d|^2 \sim \text{Exp}(\rho_d)$ and $\rho_d$ represents the amplitude of channel fading fluctuation. $\theta_{D,h_i^d}$ is the angle of departure (AoD) from the $i$-th BS to the user. Generally, the beamforming vector of this BS $\mathbf{w}_i(\theta_i)$ can be chosen from the DFT codebook with the designed phase $\theta_i$ as $\mathbf{w}_i(\theta_i) = \sqrt{\frac{1}{M_T}} \mathbf{a}_{M_T}^H (\theta_i)$.	Then the beam gain of the direct channel can be expressed as $D_i^d = |\mathbf{w}_i(\theta_i) \mathbf{a}_{M_T}(\theta_{D,h_i})|^2 = \frac{1}{M_T} \left|\mathcal{G}_1(\theta_{D,h_i},\theta_i) \right|^2$, where we have $\mathcal{G}_1(\theta_1,\theta_2) = \frac{\sin\left(\frac{M_T}{2}\left(\sin(\theta_{1})-\sin(\theta_2)\right)\right)}{\sin\left(\frac{1}{2}\left(\sin(\theta_1)-\sin(\theta_2)\right)\right)}$. Thus, the average directional beam gain with probability (w.p.) is\vspace{-1mm}
	\begin{equation}\label{Dd}\vspace{-1mm}
		D_i^d =\left\{\begin{matrix}
			M_T	& \text{w.p.} \ \frac{\arcsin(\frac{1}{M_T})}{\pi} \\
			m_T	& \text{w.p.} \ 1-\frac{\arcsin(\frac{1}{M_T})}{\pi}
		\end{matrix}\right.
	\end{equation}
	which assumes $\theta_{D,h_i^d} \sim U(-\pi,\pi)$, and $m_T$ characterizes the average antenna gain of the side lobe.

	\subsubsection{The RIS-assisted channel} Similarly, the virtual LoS channel between the $j$-th BS and the user with the reflection of the $r$-th RIS can be expressed as $\mathbf{h}_{j,r}^v \!\!=\! \chi_v(d^{B-R}_{j,r},d^{R-U}_r) \xi_{j,r}^v \mathbf{a}(\theta_{D,h_{j,r}^v}\!) \mathbf{a}^H(\theta_{A,h_{j,r}^v}\!)\boldsymbol{\Theta}_r \mathbf{a}(\theta_{D,h_{r}^v})$, where the path loss $\chi_v(d^{B-R}_{j,r},d^{R-U}_r) \!=\! C_0 \!\left( \!d^{B-R}_{j,r}\! d^{R-U}_r\! \right)^{-\alpha}\!$, and $d^{B-R}_{j,r}$ and $d^{R-U}_r$ represent the distance from the $j$-th BS to the $r$-th RIS and from the $r$-th RIS to the user, respectively. $\xi_{j,r}^v$ is the small-scale fading channel gain with $\xi_{j,r}^v \sim \text{Exp}(\rho_v)$ and $\rho_v$ characterizes the amplitude of virtual LoS channel fading fluctuation. $\theta_{D,h_{j,r}^v}$, $\theta_{A,h_{j,r}^v}$ and $\theta_{D,h_{r}^v}$ represent the AoD from the $j$-th BS, the angle of arrival (AoA) to the $r$-th RIS and the AoD from the $r$-th RIS, respectively. $\boldsymbol{\Theta}$ is the reflection matrix introduced by the $r$-th RIS. Similarly, the antenna gain can be further written as $D_j^v = D_j^d \left| \mathcal{G}_2(\theta_{A,h_{j,r}^v},\theta_{D,h_{r}^v},\phi_r) \right|^2$,	where $\mathcal{G}_2(\theta_1,\theta_2,\phi) = \frac{\sin\left(\frac{N_R}{2}\left(\sin(\theta_{1})-\sin(\theta_2)+\phi\right)\right)}{\sin\left(\frac{1}{2}\left(\sin(\theta_1)-\sin(\theta_2)+\phi\right)\right)}$. Then, the statistical model of directional beam gain of the cascaded link can be expressed as\vspace{-1mm}
	\begin{equation}\label{Dv}\vspace{-1mm}
		D_j^v =\left\{\begin{matrix}
			M_T N_R^2,	& \text{w.p.} \ \frac{\arcsin(\frac{1}{N_R})\arcsin(\frac{1}{M_T})}{\pi^2} \\
			M_T n_R^2,	& \text{w.p.} \ \frac{(\pi-\arcsin(\frac{1}{N_R}))\arcsin(\frac{1}{M_T})}{\pi^2}\\
			m_T N_R^2,  & \text{w.p.} \ \frac{(\pi-\arcsin(\frac{1}{M_T}))\arcsin(\frac{1}{N_R})}{\pi^2} \\
			m_T n_R^2, 	& \text{w.p.} \ \frac{(\pi-\arcsin(\frac{1}{M_T}))(\pi-\arcsin(\frac{1}{N_R}))}{\pi^2}
		\end{matrix}\right.
	\end{equation}
	where $\theta_{A,h_{j,r}^v}, \theta_{D,h_{r}^v} \sim U(-\pi,\pi)$ and $n_T$ represents the average antenna gain of the side-lobe at the RIS.
	\vspace{-1mm}
	\subsection{Association Policies}
	During downlink transmission, the typical user receives desired communication signals from the associated BS, referred to as the serving BS. Then, the serving BS will receive the echoes reflected by the typical user, and extract the sensing information from echo signals. The other BSs also send signals that interfere with the ISAC process of the typical user, referred to as interfering BSs.
	
	There are two possible types of serving BSs according to the association policy, which is based on the average channel gain and ignores the effect of small-scale fading. In particular, we select the BS among those capable of establishing either LoS or VLoS path that can achieve the maximum channel gain to associate with the typical user as follows, $b^d = \arg \max_{b_i \in \Phi_B^L} \chi_d(d^{B-U}_i) M_T$ and $b^v = \arg \max_{b_j \in \Phi_B^N, r \in \Phi_R^L} \chi_v(d^{B-R}_{j,r},d^{R-U}_r) M_T N_R^2$, where the maximum directional beam gain is employed at the serving BS, i.e., $M_T$ and $M_T N_R^2$. The maximum VLoS path is very complicated to obtain since it requires searching for all NLoS BSs and LoS RISs. Hence, we will ultilize the greedy algorithm with low complexity to select the nearest NLoS BS and LoS RIS respectively as $b^v = \arg \max_{b_j \in \Phi_B^N} \left(d^{B-U}_j \right)^{-\alpha}$ and $r^v = \arg \max_{r \in \Phi_R^L} \left( d^{R-U}_r \right)^{-\alpha} $.
	
	Consequently, the association policy with the $b$-th BS can be identified to choose a larger channel gain between $b^d$ and $b^v$ as follows $b = \arg \max \left( \chi_d(d^{B-U}_{b^d}) M_T , \ \chi_v(d^{B-R}_{b^v,r},d^{R-U}_r) M_T N_R^2  \right)$.
	
	\section{Coverage Rate in ISAC Networks}
	The ISAC coverage rate $P^{cs}(\epsilon_1,\epsilon_2)$ is defined as the probability that the received communication SINR $\gamma^c$ at the typical user and sensing SINR $\gamma^s$ at the serving BS are simultaneously larger than the predefined thresholds $\epsilon_1>0$ and $\epsilon_2>0$, i.e., $P^{cs}(\epsilon_1,\epsilon_2) = \mathbb{P}(\gamma^c \ge \epsilon_1, \gamma^s \ge \epsilon_2)$. Note that the ISAC dual functions share the same network topology, i.e., the identical locations of BSs and RISs. Hence, the communication and sensing performance are coupled to each other. Alternatively, we should first consider the conditional coverage rate for associating with the LoS BS and VLoS BS, respectively.   
	
	We first derive the conditional coverage rate when the typical user is associated with the LoS serving BS, where the communication and sensing SINR can respectively expressed as $\gamma_d^c = \frac{C_0 (d_{0,L}^{B-U} )^{-\alpha} |\xi_0^d|^2 M_T}{ I_{c,d}^L+I_{c,d}^V+\sigma_c^2}$ and $\gamma_d^s = \frac{P_s C_0 (d_{0,L}^{B-U} )^{-2\alpha} |\kappa|^2 M_T M_R}{ I_{s,d}^L+I_{s,d}^V+\sigma_s^2}$. The interference signal strength can be written as \small$I_{c,d}^L \!\!=\!\! \sum\limits_{b_i \in \Phi_B^L \setminus b_0} C_0 (d_i^{B-U})^{-\alpha} |\xi_i^d|^2 D_i^d$, $I_{c,d}^V =  \sum\limits_{b_j \in \Phi_B^V} C_0 (d_{j,r}^{B-R} d_r^{R-U})^{-\alpha} |\xi_{j,r}^v|^2 D_j^v$, $I_{s,d}^L = \sum\limits_{b_i \in {\Phi}_B^L \setminus b_0} C_0 (d_{i}^{B-B})^{-\alpha}\\ |\xi_i^{s,d}|^2 D_i^{d}$ and $I_{s,v}^V = \sum\limits_{b_j \in {\Phi}_B^V } C_0 (d_{j}^{B-B})^{-\alpha} |\xi_{j}^{s,d}|^2 D_j^{d}$\normalsize. $P_s$ is the post-processing power gain at the serving BS; $\kappa \sim \text{Exp}(1)$ is the radar cross section. $\sigma_c^2$ and $\sigma_s^2$ are the noise power for the communication and sensing process, respectively.
	
	These SINR calculations require first deriving the distance distribution of the serving LoS path and the cascaded path of the VLoS BS, derived in the following lemma.
	
	\begin{lemma}
		The probability density functions (PDFs) of the distance from the typical user to the nearest LoS BS and VLoS BS and the cascaded path length of the nearest NLoS BS reflected by the nearest LoS RIS are respectively given by $f_{d_{0,L}^{B-U}}(x) = 2\pi \lambda_B^L(x) x \exp\left( -2\pi \int_0^x \lambda_B^L(r) r \mathrm{d} r \right)$, $f_{d_{0,V}^{B-U}}(x) = 2\pi \lambda_B^V(x) x \exp\left( -2\pi \int_0^x \lambda_B^V(r) r \mathrm{d} r \right)$ and $f_{\eta_0}(\eta)\\ = \int_0^{\infty} f_{\eta_0 | d_r^{R-U}}(\eta | y) f_{d_r^{R-U}}(y) \mathrm{d} y$, where $f_{d_r^{R-U}}(y) = 2\pi\lambda_R^L(y) y \exp\left( -2\pi \int_0^y \lambda_R^L(r) r \mathrm{d} r \right)$, $f_{\eta_0 | d_r^{R-U}}(\eta | y) = \frac{\partial }{\partial \eta} \\ \exp(-\!\!\int_{-\pi}^{\pi}\!\! \int_{(y \cos\theta \!-\! \sqrt{-y^2 \sin^2\theta + \frac{\eta^2}{y^2}})^+}^{(y \cos\theta \!+\! \sqrt{-y^2 \sin^2\theta + \frac{\eta^2}{y^2}})^+}\!\!\! \lambda_B^V(r) r \mathrm{d} r \mathrm{d} \theta ) $ and $(x)^+ = \max(x, 0)$.
		
		\proof
		Please see Appendix~\ref{lemma1}.
		\endproof
		
	\end{lemma}

	\begin{proposition}
		The coverage rate of the ISAC function when associating with the LoS serving BS is derived as
			\begin{align}\label{p_d^cs}\vspace{-2mm}
				P_d^{cs}(\epsilon_1,\!\epsilon_2)& \!=\!\mathbb{P}(\gamma_d^c \!\ge\! \epsilon_1\!,\! \gamma_d^s \!\ge\! \epsilon_2\!)\! =\!\!  \int_0^{\infty}\!\!\!\!\! \exp\!\left(\!\! -\frac{\epsilon_1 M_R \sigma_c^2\! \!+\! \epsilon_2 x^{\alpha}\! \sigma_s^2 }{C_0 x^{-\alpha} M_T M_R} \!\right)\nonumber\\
				& \cdot \Xi_1^L(x,\epsilon_1,\epsilon_2) \Gamma_1^V(x,\epsilon_1,\epsilon_2,\mathcal{S}_1)  f_{d_{0,L}^{B-U}}(x) \mathrm{d} x
			\end{align}
		where\vspace{-2mm}\small
			\begin{align}
				&\Xi_1^L(x,\epsilon_1,\epsilon_2) \!=\! \exp \!\Bigg\{\!\! -\!\! \int_{-\pi}^{\pi}\!\! \int_{x}^{\infty}\!\! \bigg[ 1 \!-\! \sum_{k1=1}^2 \!\sum_{k2=1}^2 \frac{b_{k1}^d M_R}{M_R \!+\! \epsilon_1 (\frac{r}{x})^{-\alpha} a_{k1}^d} \nonumber\\
				&  \frac{b_{k2}^d \rho_d^s P_s M_T M_R}{\rho_d^s P_s M_T M_R + \epsilon_2 (x^2+r^2\!-\!2xr\cos\phi)^{-\frac{\alpha}{2}}\! x^{2\alpha} a_{k2}^d}\bigg] \nonumber\\
				&\cdot \lambda_B^L(\sqrt{x^2+r^2-2xr\cos\phi}) r \mathrm{d}r\mathrm{d}\phi \Bigg\},
			\end{align}
		\begin{equation}
			\begin{aligned}
				&\Gamma_1^V(x,\epsilon_1,\epsilon_2,\mathcal{S}_1)=\int_0^{\infty} f_{d_r^{R-U}}(y)  \exp \!\! \bigg[\! -\!\! \frac{1}{2\pi}\! \int_{-\pi}^{\pi}\! \int_{\mathcal{S}_1} \!\bigg(\! 1-\\
				&\! \sum_{k1=1}^4 \sum_{k2=1}^2 \frac{b_{k1}^v \rho_v M_R}{\rho_v M_R + \rho_d \epsilon_1   (\frac{y}{x})^{-\alpha} (r^2+y^2-2r y \cos \theta)^{-\frac{\alpha}{2}} a_{k1}^v} \\
				&\! \frac{b_{k2}^d \rho_d^s P_s M_T M_R}{\rho_d^s P_s\! M_T M_R \!\!+\! \epsilon_2 (x^2\!+\!r^2\!\!-\!2xr\cos\phi)^{-\frac{\alpha}{2}}\! x^{2\alpha} a_{k2}^d} 
				\!\!\bigg)\! \lambda_B^V(r) r \mathrm{d} r\mathrm{d} \theta \mathrm{d} \phi  \!\bigg] \! \mathrm{d} y,
			\end{aligned}
		\end{equation}
		$\mathcal{S}_1\!\! = \!\!\bigg\{\! (r,\theta)\!:\! r \!\in\! \bigg[0, \!\Big(\! y\!\cos\theta\!-\!\sqrt{\!-y^2 \sin^2\!\theta \!+\! x^2 N_R^{\frac{4}{\alpha}} y^{-2}} \! \Big)^+\!\! \bigg] \cup \bigg[ \big(y\cos\theta+\sqrt{-y^2 \sin^2\theta + x^2 N_R^{\frac{4}{\alpha}} y^{-2}}\big)^+, \infty  \bigg] \bigg\}$.
		\normalsize
		
		\proof
		Please see Appendix~\ref{theorem7}.
		\endproof
		
	\end{proposition}
	\vspace{-2mm}
	Then, another conditional coverage rate of associating with the VLoS serving BS can be further derived, where the communication and sensing SINR can be respectively expressed as $\gamma_v^c = \frac{C_0 \left(d_{0,r}^{B-R} d_r^{R-U}\right)^{-\alpha} |\xi_{0,r}^v|^2 M_T N_R^2}{I_{c,v}^L + I_{c,v}^V+\sigma_c^2}$ and $\gamma_v^s = \frac{P_s C_0 (d_{0,r}^{B-R} d_r^{R-U} )^{-2\alpha} |\kappa|^2 M_T M_R}{ I_{s,v}^L+I_{s,v}^V+\sigma_s^2}$. The interference signal can be written as \small $I_{c,v}^L = \sum\limits_{b_i \in \Phi_B^L} C_0 (d_i^{B-U})^{-\alpha} |\xi_i^d|^2 D_i^d$, $I_{c,v}^V \!\!= \!\!\sum\limits_{b_j \in \Phi_B^V \setminus b_0} C_0 (d_{j,r}^{B-R} d_r^{R-U})^{-\alpha} |\xi_{j,r}^v|^2 D_j^v$, $I_{s,d}^L = \sum\limits_{b_i \in {\Phi}_B^L} C_0 (d_{i}^{B-B})^{-\alpha}\\ |\xi_i^{s,d}|^2 D_i^{d}$ and $I_{s,v}^V = \sum\limits_{b_j \in {\Phi}_B^V \setminus b_0} C_0 (d_{j,r}^{B-R} d_r^{R-B})^{-\alpha} |\xi_{j}^{s,d}|^2 D_j^{d}$\normalsize.
	\vspace{-2mm}	
	\begin{proposition}
		The coverage rate of the ISAC function when associating with the VLoS serving BS is derived as 
			\begin{align}\label{p_v^cs}
				P_v^{cs}\!&(\epsilon_1,\!\epsilon_2) \!=  \mathbb{P}(\gamma_v^c \!\ge\! \epsilon_1\!,\! \gamma_v^s \!\ge\! \epsilon_2\!)\! =\!\!\! \int_0^{\infty}\!\!\!\!\! \exp\!\left( \!\!-\frac{  \epsilon_1\! M_R \!N_R^2 \sigma_c^2 \!+\! \eta^{\alpha} \sigma_s^2 }{C_0 {\eta}^{-\alpha} M_T M_R N_R^4} \right)\nonumber\\
				& \cdot \Xi_2^L(\eta,\epsilon_1,\epsilon_2,\mathcal{S}_2) \Gamma_2^V(\eta,\epsilon_1,\epsilon_2,\mathcal{S}_3)  f_{\eta_0}(\eta) \mathrm{d} \eta,
			\end{align}
		where\vspace{-2mm}
		\footnotesize
			\begin{align}
				&\Xi_2^L(\eta,\epsilon_1,\epsilon_2, \mathcal{S}_2) \!=\!\! \int_0^{\infty}\!\!\! f_{d_{0,V}^{B-U}}(x)\! \exp \!\bigg[ \!-\!\!\frac{1}{2\pi}\! \int_{-\pi}^{\pi}\!\int_{\mathcal{S}_2} \!\bigg(\! 1 \!- \sum_{k1=1}^2 \sum_{k2=1}^2 \nonumber\\
				&\!\frac{b_{k1}^d \rho_d M_R N_R^2}{\rho_d\! M_R\! N_R^2 \!\!+\!\! \rho_v\! \epsilon_1\! (\frac{r}{\eta})^{-\alpha} \!a_{k1}^d}\!  \frac{b_{k2}^d \rho_d^s P_s M_T M_R N_R^4}{\rho_d^s \!P_s\! M_T\! M_R\! N_R^4\! \!+ \!\!\epsilon_2 (x^2\!\!+\!r^2\!\!-\!\!2xr \cos\phi\!)^{\!-\frac{\alpha}{2}}\!\! {\eta}^{2\alpha}\! a_k^d}\!\bigg)\!\nonumber\\
				& \lambda_B^L(r)\mathrm{d} r \mathrm{d} \theta  \mathrm{d} \phi  \bigg]  \mathrm{d} x,
			\end{align}
		\begin{equation}
			\begin{aligned}
				&\Gamma_2^V(\eta,\epsilon_1,\epsilon_2,\mathcal{S}_3)\!=\!\!\int_0^{\infty}\!\!\!\!\int_0^{\infty}\!\!\!\!f_{d_{0,V}^{B-U}}(x)f_{d_r^{R-U}}(y) \exp \!\!\bigg[ \!\!-\!\! \int_{-\pi}^{\pi} \!\!\int_{\mathcal{S}_3} \!\!
				\bigg(\! 1\!-\\
				&\sum_{k1=1}^4 \sum_{k2=1}^2  \frac{b_{k1}^v M_R N_R^2}{ M_R N_R^2 + \epsilon_1   (\frac{y}{\eta})^{-\alpha} (r^2+y^2-2r y \cos \theta)^{-\frac{\alpha}{2}} a_{k1}^v } \\
				& \frac{b_{k2}^d \rho_d^s P_s\! M_T\! M_R N_R^4}{\rho_d^s P_s M_T M_R N_R^4\!\! + \!\epsilon_2 (x^2\!+\!r^2\!\!-\!\!2xr\!\cos\phi\!)^{-\frac{\alpha}{2}}\!\! {\eta}^{2\alpha} a_{k2}^d}\!  \bigg) \!\lambda_B^V(r) \mathrm{d} r \mathrm{d} \theta \mathrm{d}\phi \!\bigg] \! \mathrm{d} x \mathrm{d} y,
			\end{aligned}
		\end{equation}
		$\mathcal{S}_2 = \left\{ r: r \in [N_R^{-\frac{2}{\alpha}} \eta, \infty) \right\}$ and $\mathcal{S}_3\!= \!\bigg\{ \!(r,\theta)\!:\! r \in \!\bigg[ 0, \!\Big(\!\! y \!\cos\!\theta\!-\!\sqrt{\!-y^2 \sin^2\!\theta \!+\! \eta^2 y^{-2}} \! \Big)^+\! \bigg]\!\!\cup \!\bigg[ \! \!\left(\!\! y \!\cos\!\theta\!+\!\sqrt{\!-y^2 \sin^2\!\theta \!+\! \eta^2 y^{-2}} \! \right)^+\!\!,\infty \bigg]\! \bigg\}$.
		\normalsize
		
		\proof
		Please see Appendix~\ref{theorem8}.
		\endproof
		
	\end{proposition}
	
	Finally, we derive the coverage rate of the marginal ISAC performance.
	\begin{theorem}
		The coverage rate of RIS-assisted ISAC performance can be derived as\vspace{-2mm}
			\begin{align}\vspace{-2mm}
				&P^{cs}(\epsilon_1, \epsilon_2)\! =\! \bar{\zeta}_d \!\int_0^{\infty}\!\!\!\! \exp\left( -\frac{\epsilon_1 M_R \sigma_c^2 + \epsilon_2 x^{\alpha} \sigma_s^2 }{C_0 x^{-\alpha} M_T M_R} \right)\Xi_1^L(x,\epsilon_1,\epsilon_2)\nonumber\\
				&  \Gamma_1^V(x,\epsilon_1,\epsilon_2,\mathcal{S}_1) \left(\! 1 \!-\! F_{\eta_0}(N_R^{\frac{2}{\alpha}} x)\! \right) f_{d_{0,L}^{B-U}}(x) \mathrm{d} x + \bar{\zeta}_v \int_0^{\infty}\nonumber\\
				& \!\exp\!\!\left(\!\! -\frac{  \epsilon_1 M_R N_R^2 \sigma_c^2 + \eta^{\alpha} \sigma_s^2 }{C_0 {\eta}^{-\alpha} M_T M_R N_R^4}\! \right)\! \Xi_2^L(\eta,\epsilon_1,\epsilon_2,\mathcal{S}_2) \Gamma_2^V(\eta,\epsilon_1,\epsilon_2,\mathcal{S}_3)\nonumber\\
				& \left( 1- F_{d_{0,L}^{B-U}}(N_R^{-\frac{2}{\alpha}} \eta) \right) f_{\eta_0}(\eta) \mathrm{d} \eta,
			\end{align}\vspace{-0.5mm}
		which combines the conditional coverage rate of associating with the LoS BS and VLoS BS from \emph{Proposition 2} and \emph{3}, and draws on the idea of the distance-dependent thinning method, imposing a thinning weighted value on each distance from \emph{Lemma 2} to derive the marginal coverage rate.
	\end{theorem}
	\vspace{-1mm}
	\section{Numerical Results}\label{simulation}
	In this section, we provide numerical results that illustrate the coverage rate of the ISAC process in RIS-assisted mmWave networks. These results can well verify the efficacy of our derived analytical expressions, which encompass the nearest distance distribution, coverage rate and ergodic rate. We also delve into several insights regarding the impacts of network parameters, such as the blockage density and deployment densities of BSs and RISs on the joint performance of ISAC within the network, respectively.   
	
	The densities of blockages, BSs and RISs are $\lambda_L=300$ km${}^{-2}$, $\lambda_B=100$ km${}^{-2}$ and $\lambda_R=600$ km${}^{-2}$. The BS is equipped with $8$-antenna transmitter and $8$-antenna receiver, while the RIS is equipped with $256$ elements. The path-loss coefficients are set as $C_0=-30$ dB and $\alpha=3.6$. The noise power of the communication and sensing process is set as $\sigma_c^2=\sigma_s^2=-89$ dBm. The antenna gain of the side lobe is -20 dB of the main lobe. Unless otherwise mentioned, other network parameters are listed as follows: $\mathbb{E}[L]=15$ m \cite{SG_RIS1}, $\rho_d=\rho_v=\rho_d^s=1$, $P_s=20$ dB, $W=200$ MHz.

\begin{figure}[t]\vspace{-3mm}
		\centerline{\includegraphics[width=0.4\textwidth]{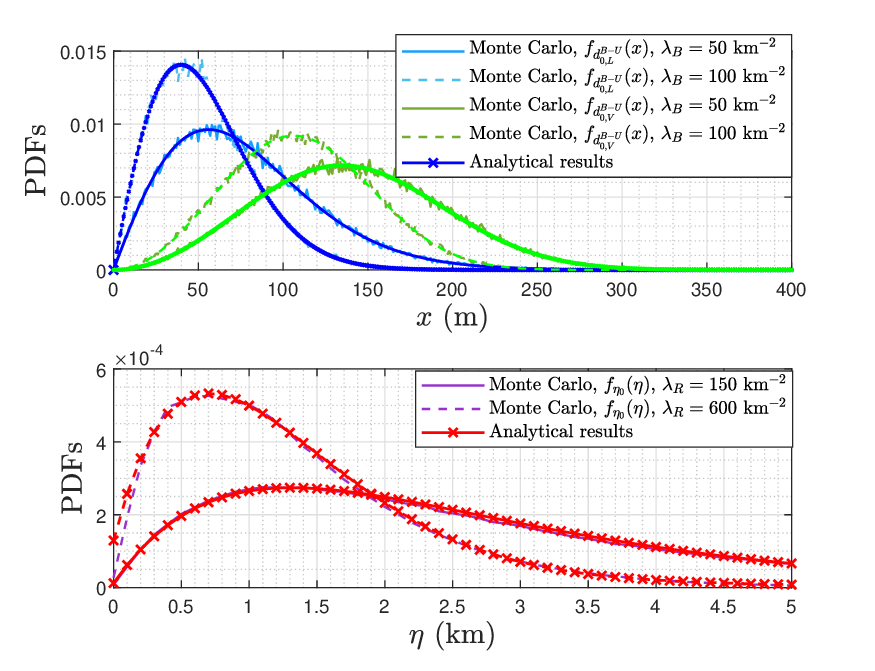}} \vspace{-2mm}
		\caption{The PDFs of $d_{0,L}^{B-U}$, $d_{0,V}^{B-U}$ and $\eta_0$, derived in \emph{Lemma 1}, versus different BS and RIS desenties.}\label{fig0}
	\end{figure}

	Fig.~\ref{fig0} presents analytical results for the distance distribution from the typical user to the nearest LoS BS, the nearest VLoS BS and the cascaded path to the nearest VLoS BS, respectively, against Monte Carlo results under different $\lambda_B$ and $\lambda_R$. It is shown that the analytical results can well overlap the experimental results, which verifies the accuracy of our derived analytical formulations. Moreover, an increase in $\lambda_B$ is associated with a decreased distance to the nearest LoS and VLoS BSs. On the other hand, increasing $\lambda_R$ can significantly reduce the path loss of the nearest VLoS BS assisted by the nearest RIS. Specifically, a fourfold increase in the density of RISs can reduce the most probable $\eta$, delivering a path loss reduction up to $-9.68$ dB.

	\vspace{-1mm}

\begin{figure}[t]\vspace{-3mm}
		\centerline{\includegraphics[width=0.4\textwidth]{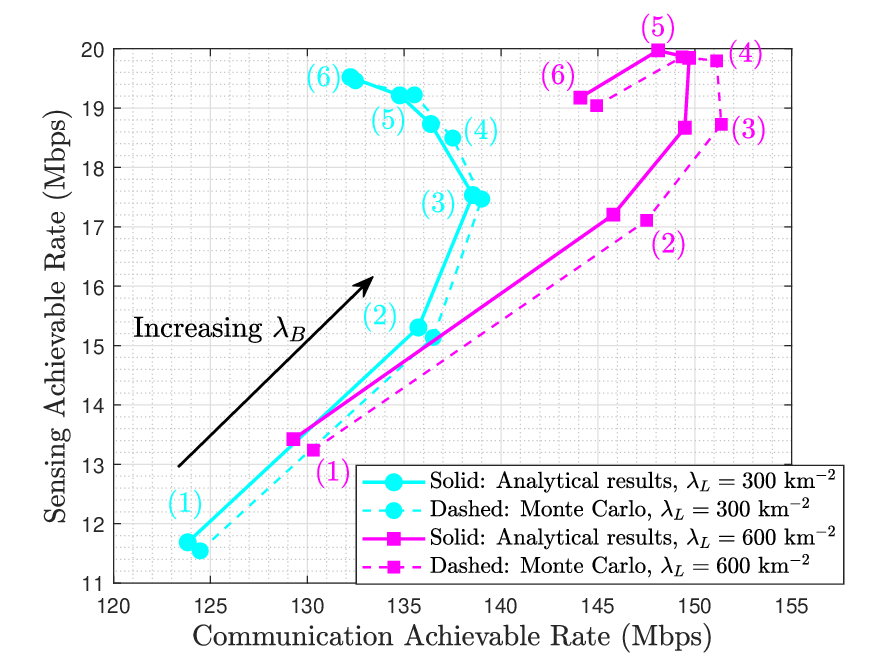}} \vspace{-2mm}
		\caption{The achievable rate trade-off between ISAC performance versus different densities of BSs, RISs and blockages.}\label{fig8}
	\end{figure}

	To delve into the effect of the blockage, BS and RIS densities on the ISAC performance jointly, we provide their tradeoff trends in Fig.~\ref{fig8}. For a fair comparison of the BS and RISs deployment, six points are selected to ensure equivalent maximum energy consumption across the network, i.e., $27$ dBw. Specifically, the energy consumption of BSs and RISs are set as $9$ dBw and $30$ dBm \cite{RIS1}, respectively, with their densities in km${}^{-2}$ as followed: $\{\lambda_B,\lambda_R\} = \{(10, 421), (20, 342), (30,262), (40, 183), (50, 104), (60,24)\}$.
	
	It can be observed from Fig.~\ref{fig8} that an increase in the BS density correlates with enhanced sensing rates, whereas the communication rate initially rises and then decreases at $\lambda_L=300$ km${}^{-2}$. Specifically, the optimal BS/RIS deployment for communication and sensing performances are identified at points $(3)$ and $(6)$, respectively. This divergence in trends is attributable to the squared path loss of sensing echoes, and thus an increase in $\lambda_B$ can significantly boost the desired sensing signal strength. However, a further increase in $\lambda_B$ adversely strengthens more interference in the communication process. At a heightened blockage density of $600$ km${}^{-2}$, the sensing performance exhibits a different trend of initial increase followed by a decline. This is attributed to the high blockage density case, where the reduction in RIS density severely compromises sensing capabilities. Specifically, the optimal BS/RIS deployment for communication performance aligns with points $(3)$ or $(4)$, while point $(5)$ is preferable for sensing performance. Consequently, the network deployment with $\lambda_B=40$ km${}^{-2}$ and $\lambda_R=183$ km${}^{-2}$ can achieve a favorable trade-off within ISAC networks.
	
\begin{figure}[t]\vspace{-3mm}
		\centerline{\includegraphics[width=0.4\textwidth]{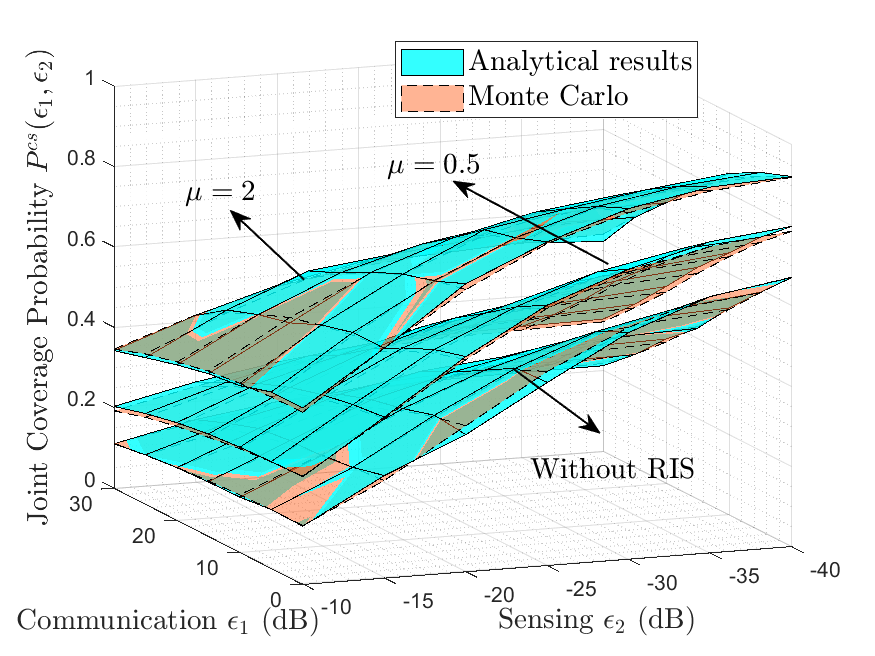}} \vspace{-2mm}
		\caption{The joint coverage rate of ISAC performance, derived in \emph{Theorem 1}, versus different SINR thresholds and RIS densities under $\lambda_L=600$ km${}^{-2}$.}\label{fig7}
	\end{figure}

	Fig.~\ref{fig7} depicts the joint ISAC coverage rate under different RIS scaling factors $\mu$ with $\lambda_R = \mu \lambda_L$. The Monte Carlo results are observed to almost overlap the analytical results, verifying the accuracy of \emph{Theorem 1}. It can provide useful insights into the network deployment strategies aimed at simultaneously ensuring the ISAC dual-function performance. Notably, the joint coverage rate exhibits a very noticeable increase with a moderate amount of RIS and can achieve $0$ dB communication SINR and $-40$ dB sensing SINR grows from $67.1\%$ to $92.2\%$ with the deployment of RISs.  
	
	
	\vspace{-1mm}
	\section{Conclusion}
	This paper established a comprehensive framework for RIS-assisted mmWave ISAC networks with the blockage effect using stochastic geometry tools. Based on this framework, the beam pattern model and user association policies were proposed to characterize the strengths of desired signals and interference boundaries on the condition of associating with the LoS BS and VLoS BS, respectively. Taking into account the inherent coupling effect of ISAC network topology, the conditional coverage rate of the ISAC performance under these two association scenarios was derived, as well as the marginal results employing the distance-dependent thinning method. All these derived theoretical results were verified accurate and demonstrated the impact of the blockage, BS and RIS densities on the ISAC performance. These findings enabled the determination of practical BS and RIS densities to achieve the tradeoff between ISAC and illustrated the joint ISAC coverage rate could rise from $67.1\%$ to $92.2\%$ with the RIS deployment.    
	
	\vspace{-2mm}
	\appendix
	\subsection{Proof of Lemma 1}\label{lemma1}
	The cumulative distribution function (CDF) of the nearest LoS BS can be covered to the non-existence of an NLoS BS within $d_{0,L}^{B-U}$ as $F_{d_{0,L}^{B-U}}(x)= 1-\exp\left( -2\pi \int_0^x \lambda_B^L(r) r \mathrm{d} r \right)$. Then, the corresponding PDF is $f_{d_{0,L}^{B-U}}(x) = 2\pi \lambda_B^L(x) x \exp\left( -2\pi \int_0^x \lambda_B^L(r) r \mathrm{d} r \right)$.
	
	Similarly, the PDF of the distance from the typical user to the nearest VLoS BS can be written as $f_{d_{0,V}^{B-U}}(x) = 2\pi \lambda_B^V(x) x \exp\left( -2\pi \int_0^x \lambda_B^V(r) r \mathrm{d} r \right)$.
	
	Let $\eta_0$ denote the path length of the VLoS link with $\eta_0=d_{0,r}^{B-R} d_r^{R-U}$, we have $F_{\eta_0}(\eta) = \int_0^{\infty} F_{\eta_0 | d_r^{R-U}}(\eta | y) f_{d_r^{R-U}}(y) \mathrm{d} y$.
	
	Since $d_r^{R-U}$ is the nearest LoS RIS from the typical user, its PDF can be obtained as $f_{d_r^{R-U}}(y) = 2\pi\lambda_R^L(y) y \exp\left( -2\pi \int_0^y \lambda_R^L(r) r \mathrm{d} r \right)$ and $F_{\eta_0 | d_r^{R-U}}(\eta | y)  = \mathbb{P} \left( y \sqrt{y^2 + (d_{0,V}^{B-U})^2 -2 y d_{0,V}^{B-U} \cos \theta} \le \eta  \right)$.

	\vspace{-3mm}
	\subsection{Proof of Lemma 2}\label{lemma2}
	The scenario where the typical user is associated with the LoS BS is that it achieves a larger channel gain than the nearest VLoS BS, i.e., $\zeta_d = \bar{\zeta}_d \cdot \mathbb{P} \left( (d_{0,L}^{B-U})^{-\alpha} \ge  \eta_0^{-\alpha} N_R^2 \right)$, where $\bar{\zeta}_d$ represents the probability of the typical user being able to connect with at least one LoS BS, i.e., $\bar{\zeta}_d = 1-\exp\left( -2\pi \int_0^{\infty} \lambda_B^L(r) r \mathrm{d} r \right)$.
	
	Similarly, the probability that the typical user is associated with the VLoS BS is written as $\zeta_v = \bar{\zeta}_v \cdot \mathbb{P} \left( (d_{0,L}^{B-U})^{-\alpha} \le  \eta_0^{-\alpha} N_R^2 \right)$, where we have $\bar{\zeta}_v = 1-\exp\left( -2\pi \int_0^{\infty} \lambda_B^V(r) r \mathrm{d} r \right)$.

	\subsection{Proof of Proposition 2}\label{theorem7}
	The ISAC coverage rate on the condition of associating with the LoS BS can be expressed as
	\small
	\begin{equation}\label{P_ISAC}
		\begin{aligned}
			&\mathbb{P}\!\left(\! \frac{C_0 (d_{0,L}^{B-U} )^{-\alpha} |\xi_0^d|^2 M_T}{ I_{c,d}^L+I_{c,d}^V+\sigma_c^2} \!\ge\! \epsilon_1, \frac{P_s C_0 (d_{0,L}^{B-U} )^{-2\alpha} |\kappa|^2 M_T M_R}{ I_{s,d}^L+I_{s,d}^V+\sigma_s^2}\! \ge \!\epsilon_2 \!\right) \\
			& \!=\! \mathbb{P}\!\! \left(\!\! |\xi_0^d|^2 \ge \frac{\epsilon_1\! (\!I_{c,d}^L\!+\! I_{c,d}^V\!+\!\sigma_c^2\!)}{C_0 (d_{0,L}^{B-U} )^{-\alpha} M_T},  |\kappa|^2\ge  \frac{\epsilon_2 (I_{s,d}^L\!+\!I_{s,d}^V\!+\!\sigma_s^2)}{P_s C_0 (d_{0,L}^{B-U}\!)^{-2\alpha}\! M_T M_R } \!\!\right)\!\!.
		\end{aligned}
	\end{equation}
	\normalsize
	Since $\xi_0^d$ and $\kappa$ are independent random variables following exponential distribution, Eq.~\eqref{P_ISAC} can be further written as
	\small
	\begin{equation}\label{theo7_1}
		\begin{aligned}
			\mathbb{E}_{x} \!\! \left[ \!\exp \!\! \left(\!\! - \frac{\rho_d \epsilon_1 (I_{c,d}^L\!+\!I_{c,d}^V\!+\!\sigma_c^2)}{C_0 x^{-\alpha} M_T} \!\!\right) \exp \!\!\left(\!\! - \frac{\epsilon_2 (I_{s,d}^L\!+\!I_{s,d}^V\!+\!\sigma_s^2)}{P_s C_0 x^{-2\alpha} M_T M_R} \!\!\right)\! \right].
		\end{aligned}
	\end{equation}
	\normalsize
	
	Depending on the type of interfering signal, we divide the expectation part inside Eq.~\eqref{theo7_1} into three parts as Eq.~\eqref{theo7_2} on the top of this page.
	\begin{figure*}[ht]
		\small
		\begin{equation}\label{theo7_2}
			\vspace{-5mm}
			\begin{aligned}
				&\exp\Bigg[ - \Bigg(\overbrace{\sum\limits_{b_i \in \Phi_B^L \setminus b_0} \rho_d \epsilon_1 M_R  (d_i^{B-U})^{-\alpha} |\xi_i^d|^2 D_i^d + \epsilon_2 P_s^{-1} x^{\alpha} (d_{i}^{B-B})^{-\alpha} |\xi_i^{s,d}|^2 D_i^{d}}^{\text{Part 1: LoS Interfering BSs}} \\
				&+  \overbrace{\sum\limits_{b_j \in \Phi_B^V} \rho_d \epsilon_1 M_R(d_{j,r}^{B-R} d_r^{R-U}\!)^{-\alpha} |\xi_{j,r}^v|^2 D_j^v + \epsilon_2 P_s^{-1} x^{\alpha}  (d_{j}^{B-B})^{-\alpha} |\xi_{j}^{s,d}|^2 D_j^{d}}^{\text{Part 2: VLoS Interfering BSs}}+ \overbrace{\frac{\epsilon_1 M_R \sigma_c^2 + \epsilon_2 x^{\alpha} \sigma_s^2}{C_0}}^{\text{Part 3: Noise}} \Bigg) \Big/ (x^{-\alpha} M_T M_R) \Bigg].
			\end{aligned}
		\end{equation}
		\normalsize
		\vspace{-5mm}
	\end{figure*}
	
	Specifically, the first part can be written as
	\small
		\begin{align}\label{theo7_3}
			&\exp \!\Bigg\{\!\! -\!\! \int_{-\pi}^{\pi}\!\! \int_{x}^{\infty}\!\! \bigg[ 1 \!-\! \mathbb{E}_{|\xi^d|^2\!, D^d,|\xi^{s,d}|^2\!, D^{s,d}}\! \Big(\! \exp \!\Big(\!\Big(\! \rho_d \epsilon_1 M_R r^{-\alpha}\! |\xi^d|^2\! D^d \nonumber\\
			& +\!\epsilon_2 P_s^{-1} x^{\alpha}  (x^2\!+\!r^2\!-\!2xr\cos\phi)^{-\frac{\alpha}{2}} |\xi^{s,d}|^2 D^{d}\Big) \!\Big/\! (- x^{-\alpha} M_T M_R\!)\Big)\! \Big)\!\bigg] \nonumber\\
			&\cdot \lambda_B^L(\sqrt{x^2+r^2-2xr\cos\phi}) r \mathrm{d}r\mathrm{d}\phi \Bigg\}.
		\end{align}
	\normalsize
	
	Then, the second part can be written as
	\small
		\begin{align}
			&\mathbb{E}_{d_r^{R-U}}\! \bigg\{\!\! \exp \!\! \bigg[\!\! -\!\! \frac{1}{2\pi}\!\! \int_{-\pi}^{\pi}\!\! \int_{\mathcal{S}_2} \!\!\!\bigg(\!\! 1\!-\! \mathbb{E}_{|\xi_{r}^v|^2, D^v,|\xi^{s,d}|^2, D^{d}}\bigg( \!\!\!\exp\!\Big(\!\Big(\!\rho_d \epsilon_1 M_R   y^{-\alpha}\! \nonumber\\
			&\!(r^2\!\!+\!\!y^2\!\!-\!\!2r y \cos \theta)^{-\frac{\alpha}{2}}\! |\xi_{r}^v|^2\! D^v \!\!\!+ \!\epsilon_2 P_s^{-1}\! x^{\alpha}\!  (x^2\!+\!r^2\!\!-\!2xr\!\cos\!\phi\!)^{-\frac{\alpha}{2}} \!|\xi^{s,d}|^2 \nonumber\\
			&D^{d} \Big)\Big/  (- x^{-\alpha} M_T^2)\Big) \bigg)\! \lambda_B^V(r) r \mathrm{d} r\mathrm{d} \theta \mathrm{d} \phi  \bigg] \bigg\}.
		\end{align}
	\normalsize

	\subsection{Proof of Proposition 3}\label{theorem8}
	The ISAC coverage rate on the condition of associating with the VLoS BS can be expressed as
	\small
	\begin{equation}\label{theo8_1}
		\begin{aligned}
			\mathbb{E}_{\eta}  \!\!\left[ \!\exp\!\!\left( \!\!- \frac{\rho_v \epsilon_1 (I_{c,v}^L\!\! + \!\!I_{c,v}^V\!\!+\!\!N_0)}{C_0 \eta^{-\alpha} M_T \!N_R^2}\!\! \right)\!\! \exp\!\!\left(\!\!- \frac{\epsilon_2 (I_{s,v}^L\!\!+\!\!I_{s,v}^V\!\!+\!\!N_0)}{P_s C_0 {\eta}^{-2\alpha} M_T\! M_R N_R^4}\!\! \right)\!\! \right]
		\end{aligned}
	\end{equation}
	\normalsize
	Depending on the type of interfering signal, we divide the expectation part inside Eq.~\eqref{theo8_1} into three parts.
	
	The first part can be written as
	\small
		\begin{align}
			&\mathbb{E}_{d_{0,V}^{B-U}} \!\!\Bigg\{\!\!\! \exp \!\!\bigg[ \!\!-\!\! \frac{1}{2\pi}\!\! \int_{-\pi}^{\pi}\!\!\int_{\mathcal{S}_3} \!\bigg(\! 1 \!- \!\mathbb{E}_{|\xi^d|^2, D^d,D^{s,d}}\! \Big(\!\! \exp\!\Big(\! \rho_v \epsilon_1 M_R N_R^2 r^{-\alpha} \nonumber\\
			&\!|\xi^d|^2 \!D^d \!+\!\epsilon_2 P_s^{-1}\! \eta^{\alpha}\! (x^2\!+\!r^2\!-\!2xr \cos\phi)^{-\frac{\alpha}{2}}\!\! D^{d}\!\Big/\! (\eta^{-\alpha}\! M_T M_R N_R^4) \! \Big)\! \Big)\!\bigg)\!\nonumber\\
			& \lambda_B^L(r)\mathrm{d} r \mathrm{d} \theta  \mathrm{d} \phi  \bigg] \Bigg\}.
		\end{align}
	\normalsize
	
	The second part can be written as
	\small
		\begin{align}
			& \mathbb{E}_{\eta}\! \bigg\{\!\! \exp \!\!\bigg[ \!\!-\!\!\! \int_{-\pi}^{\pi} \!\!\int_{\mathcal{S}_4} \!\!
			\bigg(\! 1\!-\!\mathbb{E}_{|\xi^v|^2, D^v,|\xi^{s,v}|^2,D^{s,v}} \!\bigg(\!\!\exp\Big(\! \Big(\! \rho_v \epsilon_1 M_R N_R^2 y^{-\alpha}\nonumber\\
			& \! (r^2\!\!+\!\!y^2\!\!-\!\!2r y \cos \theta)^{-\frac{\alpha}{2}}\! |\xi^v|^2 D^v \!+ \! \epsilon_2 P_s^{-1} \eta^{\alpha}(x^2\!+\!r^2\!-\!2xr\cos\phi)^{-\frac{\alpha}{2}}\nonumber\\
			&|\xi^{d}|^2 \!D^{d}\Big) \Big/ (\eta^{-\alpha} M_T M_R N_R^4)\Big) \bigg)  \bigg) \lambda_B^V(r) \mathrm{d} r \mathrm{d} \theta \mathrm{d}\phi \bigg]  \bigg\}.
		\end{align}

\end{document}